\documentclass[11pt]{article}
\usepackage{ccaption}
\usepackage{amssymb}
\usepackage{times}
\usepackage{titling}
\usepackage{graphicx}
\usepackage{amsmath}
\usepackage[version=3]{mhchem}
\usepackage{color}
\usepackage{bm}
\usepackage{lineno}
\usepackage{multirow}
\usepackage{float} 

\usepackage[paper=letterpaper,twoside,top=2.0cm,bottom=2cm,left=1.5cm,
            right=1.5cm,bindingoffset=0.0cm,voffset=0.0cm]{geometry}
            
\usepackage{bm}
\usepackage[version=3]{mhchem}
\usepackage{nicefrac}

\usepackage{color}
\usepackage{gensymb}

\definecolor{darkblue}{rgb}{0,0.02,0.45}
\usepackage[colorlinks=false,citecolor=darkblue]{hyperref}

\usepackage[all]{hypcap}
\usepackage{gensymb}

\definecolor{cream}{RGB}{222,217,201}

\clearpage

\title{\bf{Field-tunable toroidal moment in a chiral-lattice magnet}}
\author{Lei Ding$^{1,}$\thanks{These authors contributed equally to this work.}, Xianghan Xu$^{2,}$\thanksmark{1}, Harald O. Jeschke$^{3}$, Xiaojian Bai$^{1}$, Erxi Feng$^{1}$, \\
Admasu Solomon Alemayehu$^{2}$, Jaewook Kim$^{2}$, Feiting Huang$^{2}$, Qiang Zhang$^{1}$, \\ Xiaxin Ding$^{4}$, Neil Harrison$^{4}$, Vivien Zapf$^{4}$, Daniel Khomskii$^{5}$, 
\\Igor I. Mazin$^{6,}$\thanks{Email: imazin2@gmu.edu, sangc@physics.rutgers.edu, caoh@ornl.gov}, Sang-Wook Cheong$^{2,}$\thanksmark{2}, Huibo Cao$^{1,}$\thanksmark{2}\vspace{0.2cm}\\
\normalsize{\it $^1$Neutron Scattering Division, Oak Ridge National Laboratory,
Oak Ridge, TN 37831, USA}\\
\normalsize{\it  $^2$Rutgers Center for Emergent Materials and Department of Physics and Astronomy, Piscataway, NJ 08854, USA}\\
\normalsize{\it  $^3$Research Institute for Interdisciplinary Science, Okayama University, Okayama 700-8530, Japan}\\
\normalsize{\it  $^4$Los Alamos National Laboratory, Los Alamos, New Mexico 87545, USA}\\
\normalsize{\it $^5$II. Physikalisches Institut, Universit\"at zu K\"oln, Z\"ulpicher Stra\ss e 77, D-50937 K\"oln, Germany}\\
\normalsize{\it  $^6$Department of Physics and Astronomy, George Mason University, Fairfax, VA 22030, USA}\\
}

\date{}

\begin{document}

\maketitle

\baselineskip24pt

{\noindent\bf A toroidal dipole moment appears independent of the electric and magnetic dipole moment in the multipole expansion of electrodynamics. It arises naturally from vortex-like arrangements of spins. Observing and controlling spontaneous long-range orders of toroidal moments are highly promising for spintronics but remain challenging. Here we demonstrate that a vortex-like spin configuration with a staggered arrangement of toroidal moments, a ferritoroidal state, is realized in a chiral triangular-lattice magnet BaCoSiO$_4$. Upon applying a magnetic field, we observe multi-stair toroidal transitions correlating directly with metamagnetic transitions. We establish a first-principles microscopic Hamiltonian that explains both the formation of toroidal states and the metamagnetic toroidal transition as a combined effect of the magnetic frustration and the Dzyaloshinskii-Moriya interactions allowed by the crystallographic chirality in BaCoSiO$_4$.}  
\clearpage

\baselineskip20pt
\baselineskip20pt A toroidal dipole moment differs from electric and magnetic
dipoles by breaking both space inversion and time reversal symmetry. In
localized spin systems, it can be generated by a head-to-tail arrangement of
magnetic moments
\cite{Gorbatsevich1994,Ederer2007,Spaldin2008,Schmid2008,Aken2007,Cheong2007,Ginzburg1984,Dubovik1990,Foggetti2019}%
.
A uniform arrangement of toroidal moments leads to the so-called ferrotoroidal
order that is of great fundamental interest in condensed matter physics and for
potential applications in spintronics
\cite{Spaldin2008,Cheong2007,Fiebig2016,Cheong2018}. From the symmetry point
of view, toroidal moments are conventionally akin to the antisymmetric
components in the linear magnetoelectric tensor. Imposed by the symmetry
constraint, the switch of toroidal moments remains challenging because of the
requirement of a crossed magnetic and electric field in, so far, the
well-studied ferrotoroidal
materials~\cite{Ressouche2010,Zimmermann2014,Baum2013}. By introducing an
extra dipole order (magnetic or/and electric), the multiferroic coupling
provides an easy way of controlling and mutually tuning of toroidal moments through
applying only a magnetic or electric field \cite{Schmid2008}. A lattice with chirality provides
a natural environment for simultaneous existence of multi-dipole orders.

\begin{figure}[t]
\centering
\includegraphics[width=0.9\linewidth]{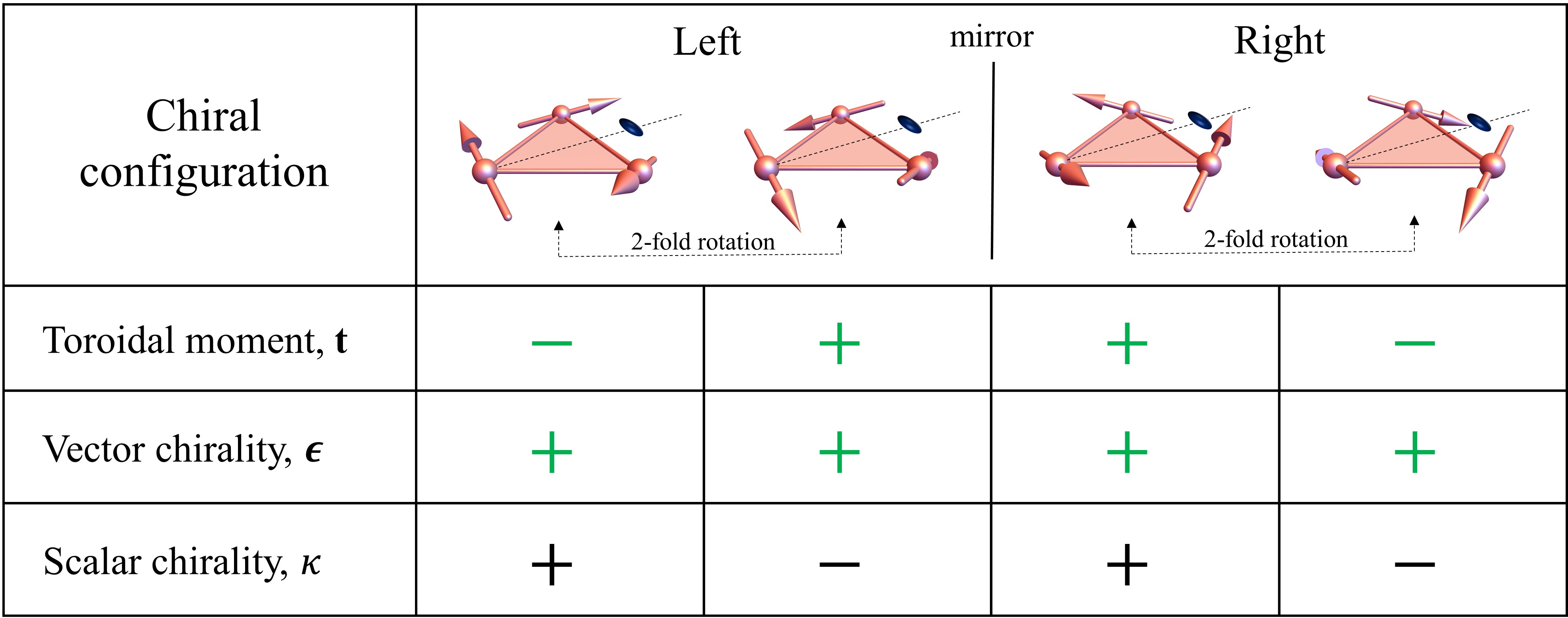}\caption{{\textbf{Toroidal
moment and magnetic chirality of spin vortex configurations with 3-fold
rotation symmetry.} Four chiral non-coplanar structures are divided into
groups of two left-handed ones and two right-handed ones. Those within each
group are connected by 2-fold rotations and those between two groups by a
mirror operation. The spins (axial vector) are represented as arrows. The dashed line
with an ellipsoid indicates one of the three 2-fold axes. Three relevant
physical quantities are defined with spins $\{\mathbf{S}_{i}=1,2,3\}$ numbered
anticlockwise: toroidal moment $\mathbf{t}\!=\!\sum_{i}\mathbf{r}_{i}%
\times\mathbf{S}_{i}$ where $\mathbf{r}_{i}$ is the vector from the center of
the triangle to spin $\mathbf{S}_{i}$; vector chirality ${\boldsymbol{\epsilon
}}\!=\!\mathbf{S}_{1}\times\mathbf{S}_{2}+\mathbf{S}_{2}\times\mathbf{S}%
_{3}+\mathbf{S}_{3}\times\mathbf{S}_{1}$; scalar chirality $\kappa
\!=\!(\mathbf{S}_{1}\times\mathbf{S}_{2})\cdot\mathbf{S}_{3}$. Green symbols
$+$ and $-$ for toroidal moment and vector chirality denote the direction of
these quantities with respect to the net magnetic moment, $+$ for parallel and
$-$ for anti-parallel. The magnetic vector chirality characterizes the sense
of spin rotation along an oriented loop (or line), while the toroidal moment
is associated with that around a center. Scalar spin chirality is a measure of
non-coplanarity which does not necessarily have a sense of rotation. In the
current example, toroidal moment and scalar chirality have a one-to-one
correspondence to the net magnetic moment for a given handedness, since all of
them are odd under time reversal. {In a crystalline material with
these chiral spin trimers as basic units, a ferro alignment of net magnetic
moments of trimers will lead to a ferro ordering of toroidal moments and
scalar chirality.}}}%
\label{fig:1}%
\end{figure}

Chirality is a geometrical property, meaning that an object subjected to the
spatial inversion cannot be superimposed upon itself by any combination of
rotations and translations \cite{Flack2003,Simonet2012}. Crystallographic chirality
has been found to be instrumental in stabilizing unusual magnetic
orders such as multiferroicity \cite{Johnson2013,Kinoshita2016}, skyrmion \cite{Muhlbauer2009,Yu2010,Seki2012}, helicity chirality \cite{Marty2008,Loire2011} and chiral magnetic soliton lattice \cite{Togawa2012}. When combined with the magnetic frustration characteristic of
antiferromagnetic interactions in equilateral triangles, spins may form a
120$^{\circ}$ vortex-like configuration \cite{Sachdev1992,Chubukov1992}, as
shown in Fig. \ref{fig:1}, generating a non-zero toroidal moment breaking both
the spatial inversion and time reversal symmetry
\cite{Sachdev1992,Chubukov1992,Batista2016}. Depending on the sense of the
in-plane spin rotations, the toroidal moment is either positive
(\textquotedblleft$+$\textquotedblright) or negative (\textquotedblleft%
$-$\textquotedblright). Manipulating toroidal moments directly by a magnetic
field is possible in a chiral magnetic vortex, if an out-of-plane spin
component is present and coupled with the in-plane spin texture through
Dzyaloshinskii-Moriya (DM) interactions \cite{DM1958,DM1960}.
Following this strategy, we find a ferritoroidal order in a unique vortex-like
spin configuration in the chiral magnet BaCoSiO$_{4}$. By applying a small
magnetic field, the toroidal moments are uniformly aligned, thereby leading to
a ferri- to ferrotoroidal transition. This toroidal transition, as well as the
simultaneously scalar ferri- to ferrochiral transition, is fully explained
within a magnetic Hamiltonian accounting for the magnetic frustration and
antisymmetric DM interactions. A key property of this Hamiltonian, as derived
from first principles calculations, is a rather special and not immediately
obvious hierarchy of Heisenberg exchange parameters, which does not correlate
with the length of the corresponding Co-Co bonds.



\begin{figure}[t]
\begin{center}
\includegraphics[width=0.95\linewidth]{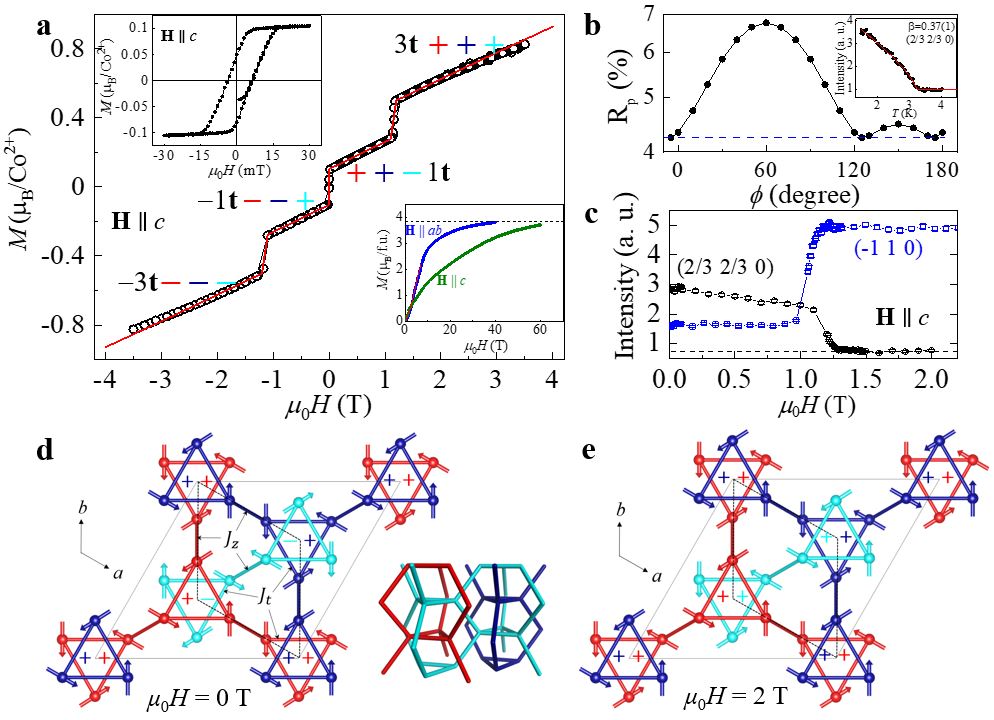}
\end{center}
\caption{\textbf{Zero-field magnetic structure and field-induced ferri- to
ferrotoroidal transition in BaCoSiO$_{4}$.} \textbf{a}, The isothermal bulk
magnetization data (open circles) measured at $2$~K with the field parallel to
the $c$ axis and the theoretical calculations (red line) from the full spin
Hamiltonian, showing a good agreement between data and calculations. The
magnetization up to $60$~T in a pulsed magnetic field at $1.5$~K is shown in
the lower inset, and the magnetization hysteresis at low fields in the upper
inset. See caption of panel \textbf{d} for the meaning of symbols $+$ and $-$.
\textbf{b}, The refined agreement factor for the powder neutron diffraction
data as a function of uniform rotation in the $ab$ plane. The dashed line
marks the best refinement. Inset shows the integrated intensity of the
magnetic reflection $(2/3\,\,2/3\,\,0)$ as a function of temperature with an
order parameter fit $\sim(1-T/T_{\text{\textrm{N}}})^{2\beta}$ (solid line),
where $\beta$ is the critical exponent. \textbf{c}, The integrated intensities
of reflection $(2/3\,\,2/3\,\,0)$ and $(-1\,\,1\,\,0)$ as a function of
magnetic field with $\mathbf{H}\parallel c$ at $1.5$ K. \textbf{d}, Zero-field
magnetic structure of BaCoSiO$_{4}$ in a $\sqrt{3}\times\sqrt{3}$ supercell
solved from powder neutron diffraction data, showing three interpenetrating
ferrotoroidal sublattices (red, blue and cyan) formed by the dominant exchange
interactions $J_{t}$ (intra-layer) and $J_{z}$ (inter-layer). The direction of
toroidal moment for each sublattice is denoted $+$ if it is parallel to the
$c$ axis and $-$ if anti-parallel. The red and blue sublattices have the same
toroidal moment, while the cyan has the opposite moment, leading to a
{ferritoroidal} state with a total moment $+1$\textbf{t}. The
primitive crystallographic unit cell is indicated by the dotted lines.
\textbf{e}, Magnetic structure of BaCoSiO$_{4}$ at $2$~T solved from
single-crystal neutron diffraction data. All spins on the cyan sublattice are
reversed, leading to a ferrotoroidal state with a total toroidal moment
$+3$\textbf{t}. {Triangles in panels \textbf{d} and \textbf{e} lie
in two adjacent layers which are bridged by the inter-layer interaction
$J_{z}$.}}%
\label{fig:2}%
\end{figure}

The stuffed tridymite BaCoSiO$_{4}$ crystallizes in the polar space group
$P6_{3}$. The crystal structure is \textit{chiral} and adopts only one
enantiomorph \cite{Liu1993}. Co atoms are tetrahedrally coordinated by oxygen with a large off-center distortion and the nearest Co atoms form spin trimers in the $ab$ plane (Fig. S1). Magnetic interactions between Co$^{2+}$ ions are
expected to be small due to long and indirect exchange paths through adjacent
SiO$_{4}$ tetrahedra. The Curie-Weiss law describes well the high-temperature
($150\lesssim T\!\lesssim\!300$~K)  magnetic susceptibility, with a negative
Weiss temperatures $\theta_{\mathrm{CW}}^{ab}\!=-10(2)$~K for $\mathbf{H}%
\parallel ab$ and $\theta_{\mathrm{CW}}^{c}\!=-26.2(4)$~K for $\mathbf{H}%
\parallel c$. The fitted Curie constants correspond to effective moments
$\mu_{\mathrm{eff}}^{ab}=4.6(4)\mu_{\mathrm{B}}$ and $\mu_{\text{\textrm{eff}%
}}^{c}=4.4(2)\mu_{\text{B}}$, consistent with the high-spin state of Co$^{2+}$
cation with $S\!=\!3/2$ and a nearly isotropic $g$-tensor of $\approx2.3$
(note that this deviates considerably from the nonrelativistic value $g=2$,
indicating sizeable spin-orbit effects). At low temperatures, the bulk
magnetic susceptibility develops an anomaly at $T_{\text{\textrm{N}}}\sim
3.2$~K, attributed to a long-range magnetic order. The isothermal
magnetization data of BaCoSiO$_{4}$ in the ordered phase are shown in
Fig.~\ref{fig:2}a, and exhibit a sequence of metamagnetic transitions.
Starting with a zero-field cooled sample, the first transition to $\sim
0.1\mu_{\text{B}}$ is observed at low fields ($\leq\!150$~Oe) for
$\mathbf{H}\parallel c$, stemming from alignment of weak ferromagnetic
domains. After a slow and linear ramp, a second transition to $\sim
0.4\mu_{\text{B}}$ occurs at a critical field $\mu_{0}H_{\text{C}}\!\sim
\!1.2$~T. This corresponds to a spin flip in one of the ferrotoroidal
sublattices, as we will elaborate further below. Similar transitions occur
with small hysteresis loops for reversed fields. Using pulsed magnetic
field, we measured the magnetization up to $60$~T along different
crystallographic directions [Fig.~\ref{fig:2}a, inset]. A much higher field is
needed to reach saturation with the magnetic field applied along the $c$ axis,
which implies the presence of a considerable easy-plane anisotropy in the $ab$
plane, formally not expected for Co$^{2+}$ in a tetrahedral environment, but
consistent with $g>2$. The slope changes around $10$ and $40$~T before the
saturation suggest additional transitions of a completely different nature.

The zero-field magnetic structure of BaCoSiO$_{4}$ was determined from powder
neutron diffraction experiments. The diffraction pattern at 1.8~K shows a set
of satellite reflections that can be indexed with a propagation vector
$\mathbf{k}=(1/3,1/3,0)$ with respect to the crystallographic unit cell. The
thermal evolution of the reflection $(2/3~2/3~0)$ confirms the
magnetic origin of the satellite reflections [Fig.~\ref{fig:2}b, inset]. A
power-law fit to the integrated intensity as a function of temperature gives a
critical exponent $\beta=0.37(1)$ and $T_{N}=3.24(1)$~K. The symmetry analysis
\cite{Bilbao} and Rietveld refinement \cite{Fullprof} yield a complex magnetic
structure where the in-plane components of magnetic moments form a vortex-like
pattern [Fig.~\ref{fig:2}d]. The structure is consistent with the magnetic
space group $P6_{3}$ (No.173.129) with a $\sqrt{3}\times\sqrt{3}$ magnetic
supercell. To test the reliability of the refined in-plane spin orientation,
we evaluate the profile factor R$_{\text{p}}$ of the fit as a function of
uniform rotation ($\phi$) of spins within the $ab$ plane. Fig.~\ref{fig:2}b
reveals clear minimums at $\phi\!\sim\!0^{\circ}$ (the structure in
Fig.~\ref{fig:2}d) and $\phi\!\sim\!120^{\circ}$, indicating the global
orientation of the spin structure with respect to the lattice is strongly
constrained. The bulk magnetization data imply the existence of a weak
ferromagnetic canting along the $c$ axis, which is allowed by the magnetic
space group symmetry, however it can not be unequivocally determined from
current neutron data. A satisfactory fit can be achieved by setting canting
angles to zero, yielding an ordered moment $m_{\text{\textrm{Co}}%
}(0\mathrm{T)}=2.71(5)\,\mu_{\text{B}}$ at $1.8$~K and $\sim\!3.67~\mu_{\text{B}}$ at zero temperature from extrapolating the power law fitting [Fig.~S2a]. 
To appreciate the toroidal
nature of the magnetic ground in BaCoSiO$_{4}$, it is instructive to decompose
the structure into three interpenetrating sublattices (red, cyan and blue)
[Fig.~\ref{fig:2}d], each of which is a network of trimers (up- and
down-triangles). Spins on every trimer form a $120^{\circ}$ configuration that
resembles a vortex and generates a non-zero toroidal moment. All trimers
belonging to a single sublattice have the identical toroidal moment, giving rise
to three ferrotoroidal sublattices. In zero field, two of them (red and blue)
have the same total moment $\mathbf{t}$, while the remaining one (cyan) has
the opposite moment, leading to a net moment of $-1\mathbf{t}$ or
$+1\mathbf{t}$ within a macroscopic magnetic domain. We dub this structure
\textit{ferritoroidal}.

The field response of the magnetic ground state in BaCoSiO$_{4}$ is
investigated using single-crystal neutron diffraction. Fig.~\ref{fig:2}c shows
the integrated intensity of the magnetic reflection $(2/3\,\,2/3\,\,0)$ and
the nuclear reflection $(-1\,\,1\,\,0)$ as a function of magnetic field along
the $c$ axis at $1.5$~K. The former is increasingly suppressed by fields and
eventually disappears at $\mu_{0}H_{\text{\textrm{C}}}\!\sim\!1.2$~T while the
latter gains significant extra intensity, indicating a field-induced
transition to a $\mathbf{k}=\mathbf{0}$ magnetic structure. The refined $\mathbf{k}%
=\mathbf{0}$ magnetic structure has the same magnetic space group symmetry as
the zero-field structure. 
The key difference is that the sublattice (cyan) with the opposite toroidal moment is flipped by $180^\circ$ in field, yielding a uniformly aligned toroidal moment for all three sublattices with a total toroidal moment $+3\textbf{t}$ [Fig.~\ref{fig:2}e], termed {\it ferrotoroidal}. This remarkable field-induced ferri- to ferrotoroidal transition directly correlates with the metamagnetic transition observed in the bulk magnetization measurements at the same critical field. 

\begin{figure}[t]
\begin{center}
\includegraphics[width=0.9\linewidth]{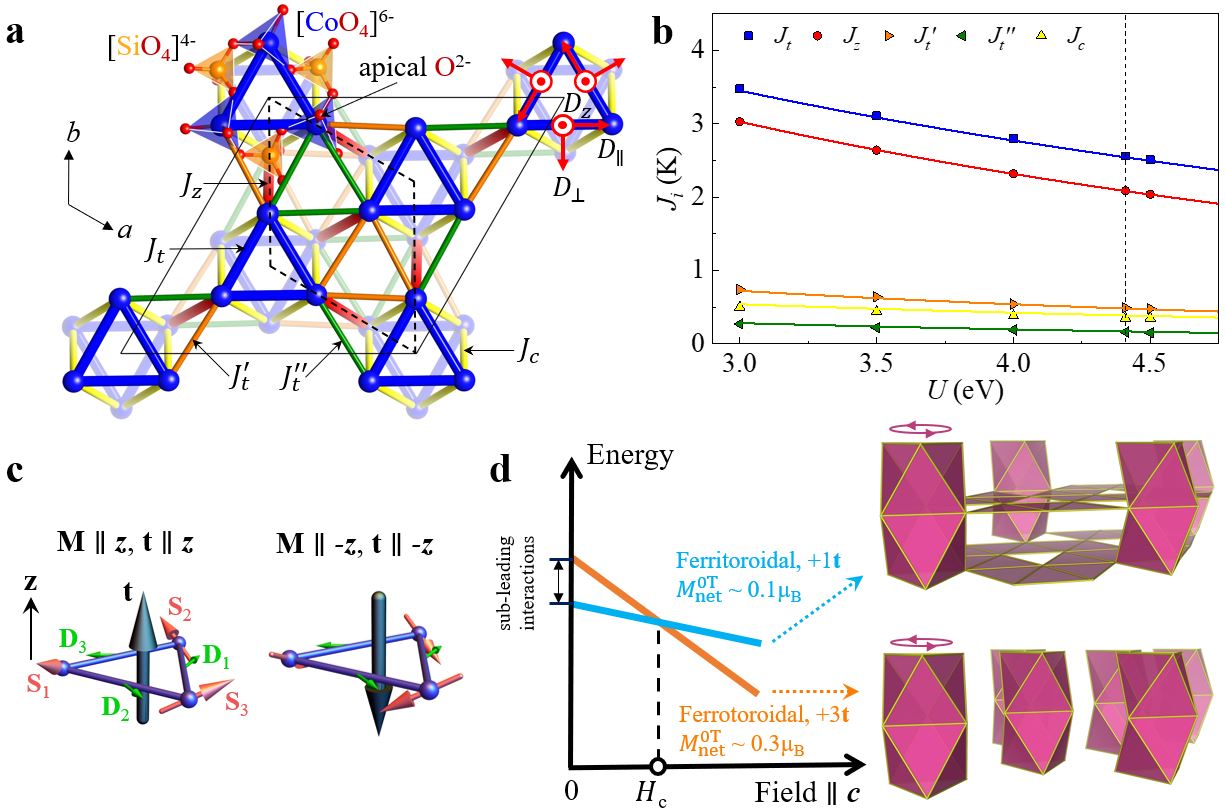}
\end{center}
\caption{\textbf{Microscopic magnetic model and the underlying mechanism for
the ferritoroidal to ferrotoroidal transition.} \textbf{a}, Magnetic exchange
pathways of BaCoSiO$_{4}$, showing three intra-layer couplings $\{J_{t},
J_{t}^{\prime}, J_{t}^{\prime\prime}\}$ and two inter-layer couplings
$\{J_{z}, J_{c}\}$. Three components of a DM vector on the nearest neighbor
$J_{t}$ bonds are indicated by red arrows in a local reference frame.
\textbf{b}, Density functional theory calculation of exchange interaction
strengths as a function of onsite interaction $U$. The dashed line marks the set of couplings with $U$ = 4.41 eV that matches the Weiss temperature from the magnetic susceptibility [Table S2]. \textbf{c}, Minimal energy
configurations for three spins $\{\mathbf{S}_{i}, i=1,2,3\}$ on a triangle
with an antiferromagnetic Heisenberg interaction and an in-plane DM
interaction. The DM vectors $\{\mathbf{D}_{i}, i=1,2,3\}$ (green arrows) are
related by 3-fold rotation symmetry and have the same sense of rotation as the
tilting of apical oxygens shown in panel \textbf{a}. { Each vector
makes a $\sim\!30^{\circ}$ angle with the bond, see main text for details.}
For this set of DM vectors, the resulting spin structure (pink arrows)
generates a toroidal moment $\mathbf{t}$ (black arrows) that is always
parallel to the magnetization $\mathbf{M}$. \textbf{d}, Energy balance between
the ferri- to ferrotoroidal state in magnetic fields. The ``frustrated''
triangular units that do \textit{not} have $120^{\circ}$ configurations (and
cost more energy) are highlighted in pink for both states. {The ferrotoroidal
state has less colored triangles in the $ab$ plane, therefore it is
energetically favored by the interactions $\{J_{t}^{\prime}, J_{t}%
^{\prime\prime}\}$, similarly the ferritoroidal state is favored by the
interaction $J_{c}$.} Competition between these subleading interactions
{results in the ferritoroidal structure with a lower energy in zero field.}
The transition to the ferrotoroidal state occurs when the energy difference is
compensated by the Zeeman energy in magnetic fields. }%
\label{fig:3}
\end{figure}

While the magnetic patterns, at first glance, seem nearly incomprehensibly
complicated, it actually has a straightforward microscopic explanation.
To show that, we first calculated the isotropic Heisenberg
magnetic interactions using the density functional theory (DFT). Inspecting
the Co $t_{2g}$ bands crossing the Fermi level, we found that the five
shortest Co-Co bonds, with $d_{\text{\textrm{Co-Co}}}$ between $5.11$ and
$5.41$ \AA , all have hopping integrals of roughly the same order [Fig.~S3] so
they should be included in the minimal model [Fig.~\ref{fig:3}a]. Next, we
performed total energy calculations for selected spin configurations and
determined exchange parameters $J$ by fitting the results to the mean field
energies of a Heisenberg model. Fig.~\ref{fig:3}b shows the fitted model
parameters as a function of onsite interaction $U$, using the room-temperature
crystal structure as input in DFT calculations. All five exchange couplings
are antiferromagnetic. Most interestingly, despite all bond lengths being
similar, two interactions stand out as dominant, independent of $U$: the
intra-layer coupling $J_{t}$ and the inter-layer coupling $J_{z}$. The Co
atoms connected by these two bonds form the three interpenetrating sublattices
shown in Fig.~\ref{fig:2}d. Within each sublattice, we expect that frustrated
$J_{t}$ interactions impose $120^{\circ}$ spin configurations on trimers that
are antiferromagnetically coupled by $J_{z}$. This explains the major
part of the experimentally determined magnetic structure.
In
addition,
relativistic DFT calculations indicate a strong easy-plane single-ion
anisotropy ($\sim\!2$~K, comparable to the dominant exchange interactions). This
is nontrivial, since Co$^{2+}$ in a tetrahedral environment features a full
$e_{g}$ shell and half-filled $t_{2g}$, and formally is not supposed to have a sizeable orbital moments.
The key is that the CoO$_{4}$ tetrahedra are considerably distorted and
moreover Co$^{2+}$ is strongly off-centered (Fig.~\ref{fig:3}a), so that the $e_{g}$
and $t_{2g}$ orbitals are actually mixed. This is also consistent with the pulsed-field
magnetization measurements. 

Of the three sublattices in Fig.~\ref{fig:2}d one (cyan) has its toroidal
moment opposite to the others. This results from the subleading exchange
interactions $\{J_{t}^{\prime},J_{t}^{\prime\prime},J_{c}\}$ [Fig.~\ref{fig:3}%
a]. The first two connect the trimers within the $ab$ plane, while the last
one connects those along the $c$ axis. A close inspection of the lattice
connectivity reveals that all three subleading interactions with the help of
$J_{t}$ form various triangular units. The total energy associated with
subleading interactions is the lowest if spins on all these triangular units
have $120^{\circ}$ arrangements (neglecting the weak ferromagnetic canting).
Yet, this condition cannot be satisfied. Fig.~\ref{fig:3}d colors all
\textquotedblleft frustrated\textquotedblright\ triangles that do \textit{not}
have $120^{\circ}$ configurations in a ferri- and ferrotoroidal state. We see
clear switching of colored triangles from one state to another, indicating a
direct competition between the intra-layer $J_{t}^{\prime}$ and $J_{t}%
^{\prime\prime}$ couplings and the inter-layer $J_{c}$ couplings. In
BaCoSiO$_{4}$, this competition favors a ferritoroidal state by a small margin
of energy in zero field.

However, this Heisenberg model, with (or without) the single-ion anisotropy,
is insufficient in explaining the weak ferromagnetic canting and the
spin-space anisotropy evidenced from our experimental data. To this end, we
introduce the antisymmetric Dzyaloshinskii-Moriya (DM) interactions
\cite{Batista2016,DM1958,DM1960} within the structural trimers
[Fig.~\ref{fig:3}a]. All three components of a DM vector on a nearest neighbor
bond are allowed due to lack of symmetry constraints. It is rather
difficult to calculate the DMI from the first principles, so for the moment we
are assuming that all three componets are present.  The DM vectors on
different bonds of the trimer are related by the 3-fold rotations. Assuming a
uniform canting along $\mathbf{c}$ and $120^{\circ}$ configurations in the
$ab$ plane for a trimer, the out-of-plane DM component $D_{z}\hat{\mathbf{z}}$
contributes $-\frac{3\sqrt{3}}{2}|D_{z}|S_{xy}^{2}$ in energy, where $S_{xy}$
is the length of in-plane spin component. Therefore, this term always favors
coplanar spin configurations {instead of} canting. Taking into account that
the $120^{\circ}$ configuration has a 3-fold vortex symmetry, we find that the
energy associated with the in-plane DM component, $\mathbf{D}_{xy},$ is
$3\sqrt{3}\left( \mathbf{D}_{xy}\cdot\mathbf{S}_{xy}\right)  \left(
\mathbf{S}_{z}\cdot\mathbf{\hat{z}}\right)  ,$ where $\mathbf{S}_{z}$ is the
out-of-plane spin component. There are two important ramifications. First, the
two components of spins are locked together, namely if $\mathbf{S}_{z}$ flips,
so does $\mathbf{S}_{xy}$ as well, to keep the DM energy gain. This is actually the essential physical factor which couples the net component of magnetization $M_z$ to the toroidal moment and allows to control toroidal moments by altering $M_z$, e.g., by external fields.
Second, to minimize the DM energy for a fixed spin canting, $\mathbf{S}_{xy}$
has to be anti-parallel to $\mathbf{D}_{xy}$ if $\mathbf{S}_{z}\parallel
\mathbf{\hat{z}}$ or parallel to $\mathbf{D}_{xy}$ if $\mathbf{S}_{z}%
\parallel-\mathbf{\hat{z}}$, which means that we can read off the direction of the
DM vector directly from the experimental spin structure, assuming $D_{z}=0$.
{In Fig.~\ref{fig:3}c, we show by green arrows the total DM vector
for each bond, which is the vector sum of $D_{\perp}$ and $D_{\parallel}$
components and makes a $\sim\!30^{\circ}$ angle with the bond, as determined
from the magnetic structure.} For each sublattice connected by $J_{t}$ and
$J_{z}$ bonds, the DM interaction creates a uniform $c$ axis canting and
corresponding in-plane toroidal spin texture. However, different sublattices
are independent of each other, hence ferri- and ferrotoroidal states would have
had the same energy, if not for the subleading Heisenberg interactions
$\{J_{t}^{\prime},J_{t}^{\prime\prime},J_{c}\}$.

We now fully understand the metamagnetic ferrotoroidal transition. Indeed,
because of the DMI-induced ferromagnetic canting, the small $\{J_{t}^{\prime
},J_{t}^{\prime\prime},J_{c}\}$-driven energy gain associated with the
ferritoroidal arrangement competes directly with the Zeeman interaction
favoring the ferrotoroidal phase. This leads to the spin-flip (also a toroidal
flip) manifested through the sudden increase of magnetization along the $c$
axis by a factor of three at the metamagnetic transition [Fig.~\ref{fig:2}e].
A direct numerical calculation of magnetization using the complete model with
all interactions discussed thus far is given in Fig.~\ref{fig:2}a and shows
good agreement with the data. See Supplementary Information and Method section
for more details and model parameters. 

At a fundamental level, the physics emerging in BaCoSiO$_{4}$
originates from its chiral crystal structure. For a single triangle in $3$ dimensions, there is a set of three mirror planes perpendicular to the
triangle and one extra mirror plane containing the triangle. The former allows
both $D_{z}$ and $D_{\perp}$ components of the DM interaction, while the
latter  allows only $D_{z}$. When $D_{\perp}$ is absent, $D_{z}$ with the
correct sign could create a vortex configuration, however the ferromagnetic
canting is unfavored in this case. Thus, the intimate coupling between
magnetization and toroidal moment is lost. In BaCoSiO$_{4}$, the triangles are
decorated by distorted { CoO$_{4}$ tetrahedra} which break all the
mirror planes while still preserving the $3$-fold symmetry. The $D_{\parallel}$
term is { then} allowed, and plays an essential role in generating
chiral vortex structures wherein the $c$ axis canting is coupled with the
sense of the spin rotation in the $ab$ plane. A direct control of toroidal
moments using only magnetic fields is therefore possible through controlling
the bulk magnetization by a uniform magnetic field, instead of using a
conjugate field such as the curl of magnetic fields. The same arguments show that at this transition the scalar chirality  $\kappa$ (Fig. 1), existing in Co triangles, also changes from the ferrichiral to ferrochiral state, similar to the three-sublattice description for the toroidal transition. The ferrochiral state with a noncoplanar spin configuration acquires the considerable Berry curvature which can lead to a variety of exotic physical phenomena such as topological magnon excitations~\cite{Katsura2010,Onose2010}.

In summary, we studied a rare chiral triangular-lattice magnet BaCoSiO$_{4}$
through bulk magnetization measurements and neutron diffraction experiments.
We uncovered a novel vortex-like spin texture and a magnetic-field-induced
ferri- to ferrotoroidal transition for the first time. Combining \textit{ab
initio} density functional theory calculations and theoretical modeling, we
have derived the microscopic energy balance and were able to
explain quantitatively the complex magnetic structure and the field-induced
metamagnetic/toroidal phase transition, neither of which had been observed before in any compound.  
 Our work shows that BaCoSiO$_{4}$ is an excellent
platform to study field-tunable toroidal moments and to explore their
interplay with the structural and magnetic chirality. Further studies on the
magnetoelectric effects and dynamical responses of toroidal spin textures are
liable to bring up further new physics and potential applications.

\newpage

\clearpage

\clearpage

\clearpage

\bibliographystyle{naturemag}

\clearpage
\subsection*{Methods}
\baselineskip20pt
\noindent\textit{Sample preparation and characterization.} 
Powder sample of BaCoSiO$_4$ was prepared by direct solid-state reaction from stoichiometric mixtures of BaCO$_3$, Co$_3$O$_4$ and SiO$_2$ powders all from (Alfa Aesar, 99.99\%) as previously reported \cite{Liu1993}. The mixture was calcined at 900 $^{\circ}$C in air for 12 hours and then re-grounded, pelletized and heated at 1200 $^{\circ}$C for 20 h and at 1250 $^{\circ}$C for 15 h with intermediate grindings to ensure a total reaction. The resulting powder sample is fine and bright blue in color. Large single crystals were grown using a laser-diode heated floating zone technique. The optimal growth conditions were growth speed of 2-4 mm/h, atmospheric air flow of 0.1 L min$^{-1}$ and counter-rotation of the feed and seed rods at 15 and 30 rpm, respectively. 
\\

\noindent\textit{Magnetization measurement.}
Temperature dependence of magnetization $M(T)$ was measured under a field of 0.1 T in a commercial magnetic property measurement system (MPMS-XL7, Quantum Design). Magnetic hysteresis loops with the field along the $a$ and $c$ axes were measured at 2 K. Magnetization up to 60 T was measured by an induction magnetometry technique \cite{Detwiler2000} using a capacitor-bank-driven pulsed magnet at the National High Magnetic Field Laboratory pulsed-field facility at Los Alamos. The pulsed-field magnetization values are calibrated against measurements in a 7 T dc magnet using a superconducting quantum interference device magnetometer (MPMS-XL7, Quantum Design). 
\\

\noindent\textit{Neutron diffraction.}
Neutron powder diffraction (NPD) experiments were performed on the time-of-flight (TOF) powder diffractometer POWGEN at the Spallation Neutron Source (SNS) at Oak Ridge National Laboratory (ORNL). A powder sample of $\sim$ 2 g was loaded in a vanadium cylinder can and measured in the temperature range of 1.8-10~K with neutron wavelength band centered at $\lambda$ =1.5~\AA~ and 2.665~\AA, covering the $d$-space range 0.5-9.0 and 1.1-15.4~\AA, respectively. Single crystal neutron diffraction experiments were carried out on the single crystal neutron diffractometer HB-3A DEMAND equipped with a 2D detector at the High Flux Isotope Reactor (HFIR), ORNL. The measurement used the neutron wavelength of 1.553~\AA~selected by a bent perfect Si-220 monochromator \cite{HB3A,Demand}. The single crystal ($\sim$0.2 g) was mounted in a vertical field superconducting cryomagnet with magnetic field up to 5.5 T and measured over the temperature range of 1.5-10 K with magnetic field applied along the $c$ axis. The data refinements were performed by FULLPROF SUITE program \cite{Fullprof}. \\

\noindent{\it Spin-polarized Density Functional Theory calculations.} Electronic structure calculations were performed using the full potential local orbital (fplo) basis~\cite{fplo} and generalized gradient approximation exchange and correlation functional~\cite{Perdew1996}. The crystal structure from Ref.~\cite{Liu1993} was used. We correct for the strong electronic correlations on Co $3d$ orbitals using a DFT+U method~\cite{Liechtenstein95} with a fixed value of the Hund's rule coupling $J_{\rm H}=0.84$~eV as suggested in Ref.~\cite{Mizokawa1996}. The Heisenberg Hamiltonian parameters were determined by an energy mapping technique~\cite{Ghosh2019}.\\

\noindent\textit{Full spin model calculations.}
The experimental magnetization data shown in Fig.~\ref{fig:2}a was modeled using a full spin Hamiltonian, including exchange interactions, single-ion anisotropy and external fields, $\mathcal{H} = \frac{1}{2}\sum_{ij}J_{ij}{\bf S}_i\cdot{\bf S}_j+A\sum_i(S_i^z)^2-g\mu_\text{B}\mu_0\sum_i{\bf H}\cdot {\bf S}_i$ with $g=2$ and $S=3/2$. Starting with the Heisenberg interactions strengths produced by DFT calculations, by trial and error, we found following set of parameters reasonably reproducing the experimental magnetization data, $J_t=2.22$~K, $J_z=1.87$~K, $J_t'=0.21$~K, $J_t''=0.61$~K, $J_c=0.48$~K, $D_\parallel=1.71$~K, $D_\perp = D_\parallel/\sqrt{3}$, $D_z=0$~K and $A=7.5$~K. Direct numerical minimization of the classical energy was performed for this model using a $\sqrt{3}\times\sqrt{3}$ supercell. The magnetization is obtained by averaging all spins of the lowest energy configuration at each field value, shown as the red line in Fig.~\ref{fig:2}a.\\

\subsection*{Acknowledgments}
\baselineskip20pt
The work at Oak Ridge National Laboratory (ORNL) was supported by the U.S. Department of Energy (DOE), Office of Science, Office of Basic Energy Sciences, Early Career Research Program Award KC0402010, under Contract DE-AC05-00OR22725. This research used resources at the High Flux Isotope Reactor and the Spallation Neutron Source, the DOE Office of Science User Facility operated by ORNL. {The work at Rutgers University was supported by the DOE under Grant No.~DOE: DE-FG02-07ER46382. The work of D. Kh. was funded by the Deutsche Forschungsgemeinschaft (DFG, German Research Foundation) - Project number 277146847 - CRC 1238. I.I.M. acknowledges support from DOE under Grant No.~DE-SC0021089. The National High Magnetic Field Laboratory is funded by the U.S. National Science Foundation through Cooperative Grant No. DMR-1157490, the U.S. DOE and the State of Florida.}  
\\
\subsection*{Author contributions}
\baselineskip20pt
S.-W.C. conceived the BaCoSiO$_4$ project. H.B.C., S.-W.C. and I.I.M. supervised this work. X.X. and A.S.A. grew the sample. X.X., J.K., F.H., X.D., N.H., V.Z. and S.-W.C. measured bulk magnetization data. L.D., E.F., Q.Z., and H.B.C. performed neutron diffraction experiments and data analysis. H.O.J., X.B., D.K. and I.I.M. performed DFT calculations and theoretical modeling. L.D., X.B., I.I.M., S.-W.C. and H.B.C. wrote the paper with comments from all the authors. \\

\subsection*{Competing financial interests}
The authors declare no competing financial interests.

\end{document}


\maketitle	
\tableofcontents
\clearpage

\section{Crystal structure and magnetic structure determination of BaCoSiO$_4$}

\begin{figure}[h]
\centering
\includegraphics[width=0.9\linewidth]{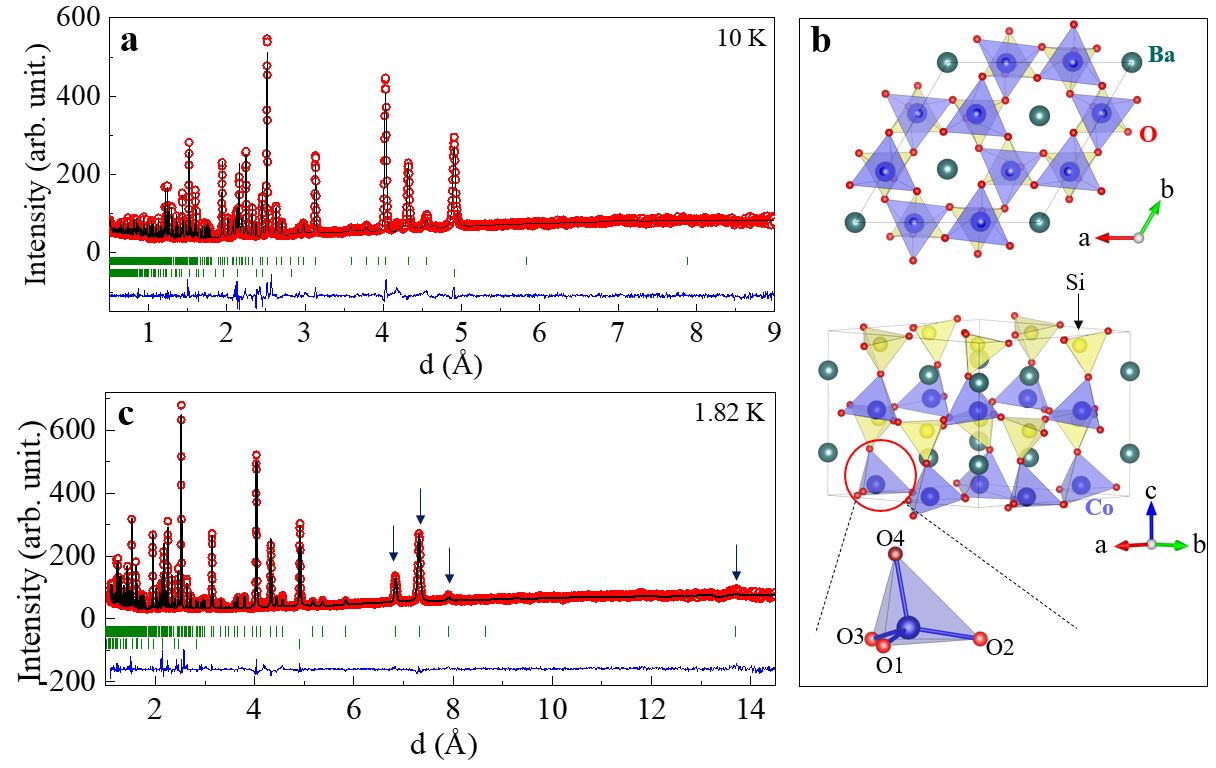}
\caption{\textbf{a}, Rietveld refinement of neutron diffraction data collected at $10$~K. Open red circles and solid black line represent experimental and calculated intensities, respectively. Solid blue line at the bottom of the panel shows the difference between them. The upper green tick marks stand for the positions of the Bragg reflections while the lower marks denote the impurity phase CoO$_3$ (weight fraction: $8.59(5)$\%). \textbf{b}, Schematic representation of the crystal structure. The distorted CoO$_4$ tetrahedron is emphasized by demonstrating the considerable off-center of the Co atoms. \textbf{c}, Rietveld refinement of neutron diffraction data collected at $1.8$~K using the magnetic space group $P6_3$. The arrows mark the magnetic reflections.} \label{fig:S1}
\end{figure}

\begin{figure}[h]
\centering
\includegraphics[width=0.9\linewidth]{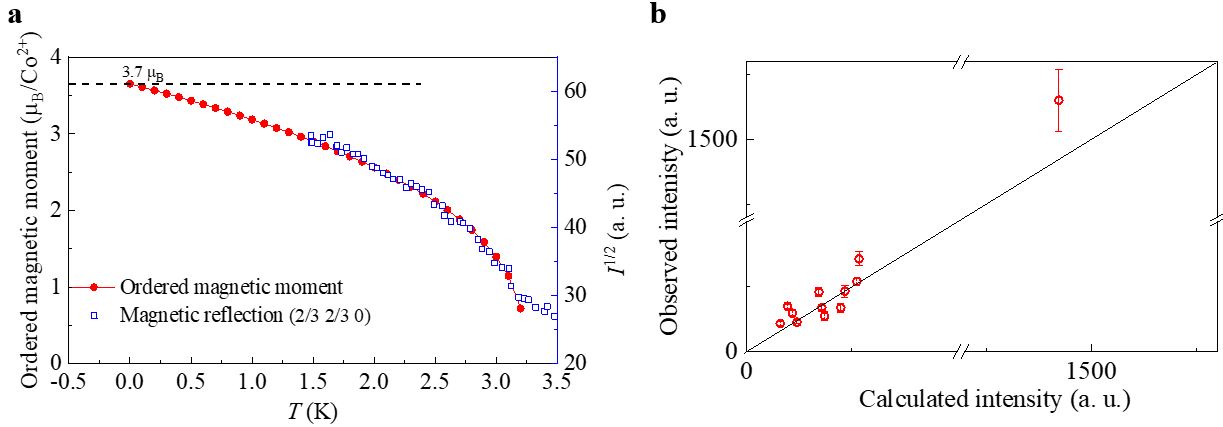}
\caption{\textbf{a}, The square root of the intensity of magnetic reflection $(2/3~2/3~0)$ and the calculated ordered magnetic moment as a function of temperature.  \textbf{b}, Observed and calculated intensity of the single crystal neutron diffraction at $1.5$~K under $2$~T.} \label{fig:S2}
\end{figure}

The structure of BaCoSiO$_4$ measured by neutron diffraction at 10 K above $T_N$ has the space group $P6_3$, consistent with the reported one from the single crystal X-ray diffraction at room temperature \cite{Liu1993}. The Rietveld refinement here adopts the reported crystal structure as a starting structural model \cite{Liu1993}. Background was described by linear interpolation of selected points in the pattern. The profile function of the powder time-of-flight neutron diffraction data was described by a convolution of a pseudo-Voigt function with a pair of back-to-back exponential, implemented by the FULLPROF SUITE program. To reduce the number of structural parameters, atoms of the same element were constrained to have the same isotropic displacement parameters. The Rietveld refinement of the neutron diffraction pattern is shown in Fig. S\ref{fig:S1}a and the corresponding structural parameters are tabulated in Table S\ref{TableS1}. As shown in Fig. S\ref{fig:S1}b, Co$^{2+}$ ions are tetrahedrally coordinated by oxygen forming spin trimers in the  $ab$ plane, which are bridged by the adjacent SiO$_4$ tetrahedra. The spin trimer layer is stacked alternately with the SiO$_4$ tetrahedra layers along the $c$ axis. The CoO$_4$ tetrahedron is considerably distorted with a large off-center of Co atoms and one relatively distant (Co-O4) bond (d$_{\text{Co-O4}}$ = 1.994~\AA) and three smaller bond lengths (d$_{\text{Co-O1}}$ = 1.946~\AA, d$_{\text{Co-O2}}$ = 1.944~\AA, d$_{\text{Co-O3}}$=1.938~\AA).  

\begin{table}[htb] 
\caption{The structure parameters of BaCoSiO$_4$ measured at 10 K by powder neutron diffraction. The space group is $P6_3$, $a = 9.1124(1)$ \AA,  $b=9.1124(1)$ \AA, $c = 8.6447(2)$ \AA, $\alpha=90^\circ$, $\beta=90^\circ$, $\gamma=120^\circ$. $R_\text{p}=4.62$\%. $R_{\text{Bragg}}= 10.5$\%. The atomic displacement parameter $B_{iso}$ is in 1/(8$\pi^2$) \AA$^2$.}\label{TableS1}
\begin{center}
\begin{tabular}{cccccc}
\hline
\hline
atom & type & $x$& $y$ &  $z$  & $B_{iso}$ \\
\hline
Ba1 & Ba  &  0 & 0 & 0.250 & 0.23(4) \\
Ba2 & Ba  &  1/3 & 2/3 & 0.224(1) & 0.23(4) \\
Ba3 & Ba  &  2/3 & 1/3 & 0.224(1) & 0.23(4) \\
Co1 & Co  &  0.681(1) & 0.672(2) &  0.529(2 & 0.39(5) \\
Si1 & Si   &  0.6576(9) & -0.0091(8)& 0.433(1) & 0.39(5)\\
O1  & O    &  0.7664(7) & 0.9146(7) &  0.524(1) & 0.56(2)\\
O2  & O    &  0.4648(6) & 0.9043(7) &  0.489(1) & 0.56(2)\\
O3  & O    &  0.7616(7) & 0.1962(7) &  0.448(1) & 0.56(2)\\
O4  & O    &  0.7251(5) & 0.6542(5) &  0.752(1) & 0.56(2)\\
\hline
\hline
\end{tabular}
\end{center}
\end{table}

Neutron powder diffraction data collected at $1.8$~K using the central wavelength of $2.665$~\AA~ were used to determine the magnetic structure without the magnetic field. The magnetic symmetry analysis was performed using the MAXMAGN tool at the Bilbao Crystallographic Server. For a given propagation vector $\textbf{k} = (1/3, 1/3, 0)$ and the parent grey group $P6_3$1$'$, there are only three k-maximal magnetic subgroups ($P6'_3$, $P6_3$, $P3$) which were tested by comparing the neutron data at $1.8$~K. We found the magnetic space group $P6_3$ (No.173.129) that is compatible with a $\sqrt{3}\times \sqrt{3}$ supercell provides a satisfactory result with the agreement factors  $R_\text{p} = 5.66$\% and $R_{\text{Bragg}} = 18.1$\%. The magnetic model as well as the nuclear phase were refined using the FULLPROF SUITE program. The best refinement and the corresponding magnetic configuration are shown in Fig. S\ref{fig:S1}c and Fig.~2d, respectively. The power-law fit of the intensity of magnetic reflection (2/3~2/3~0) (see main text) in the measured temperature range gives a way to extrapolate the intensity at zero temperature. Using the refined magnetic moment at 1.8 K, we can calculate the magnetic moment down to zero temperature. The ordered magnetic moment as a function of temperature is shown in Fig. S2a.

The determination of the magnetic structure under $2$~T was done based on the single crystal neutron diffraction data at $1.5$~K with the field along the $c$ axis using a similar procedure. Given the propagation vector \textbf{k} = $\textbf{0}$ and the parent grey space group, we found two k-maximal magnetic subgroups: $P6'_3 $ and $P6_3$. Due to the symmetry restriction, the former does not support a ferromagnetic component along the $c$ axis, and can thus be excluded. The magnetic space group $P6_3$ is compatible with the neutron diffraction data. Fig.~S\ref{fig:S2}b shows the comparison between the observed and calculated neutron diffraction reflections. The best refinement gives the agreement factors $\chi^2 = 16.4$ and $R_\text{F} = 10.5\%$ The corresponding schematic drawing of the magnetic structure is shown in Fig.~2e.

\clearpage
\section{Magnetization characterization}

\begin{figure}[h]
\centering
\includegraphics[width=0.9\linewidth]{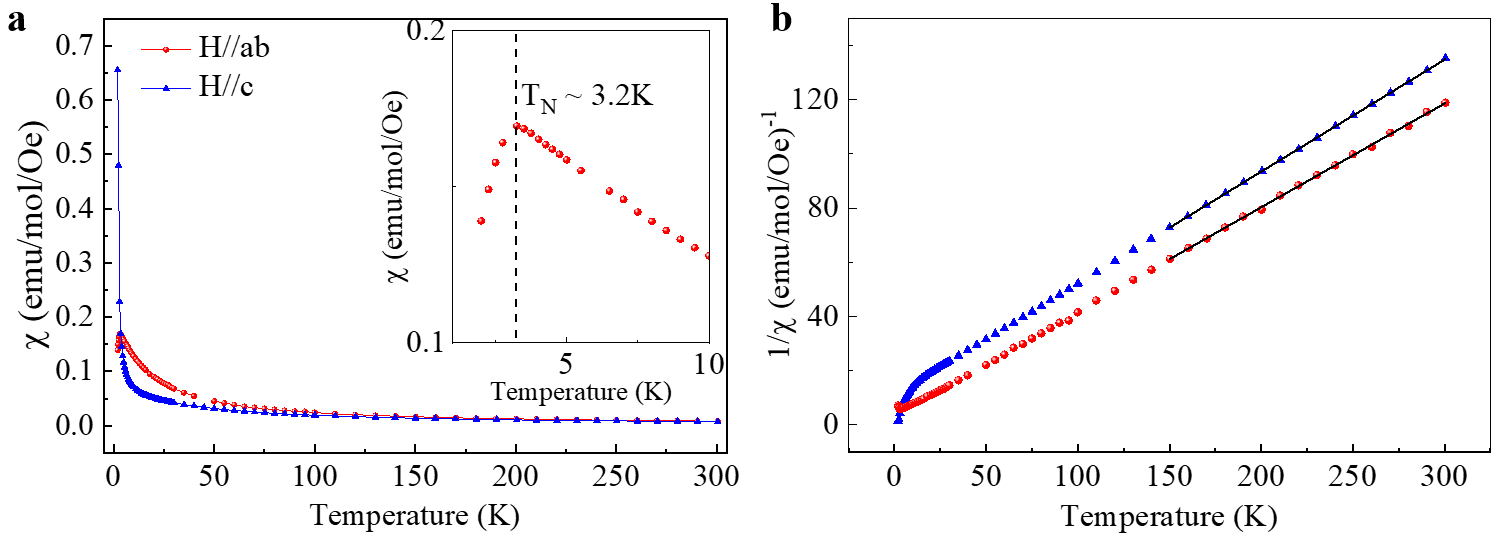}
\caption{\textbf{a}, Temperature dependence of the magnetic susceptibility of BaCoSiO$_4$ with the magnetic field ($\mu_0H= 0.1$~T) along the $c$ axis and in the $ab$ plane. Inset shows the magnetic ordering temperature. \textbf{b}, Inverse magnetic susceptibility curves and the corresponding fits using the Curie-Weiss law.} \label{fig:S3}
\end{figure}
\clearpage

\section{Density functional theory calculations}

\begin{figure*}[htb]
\includegraphics[width=\textwidth]{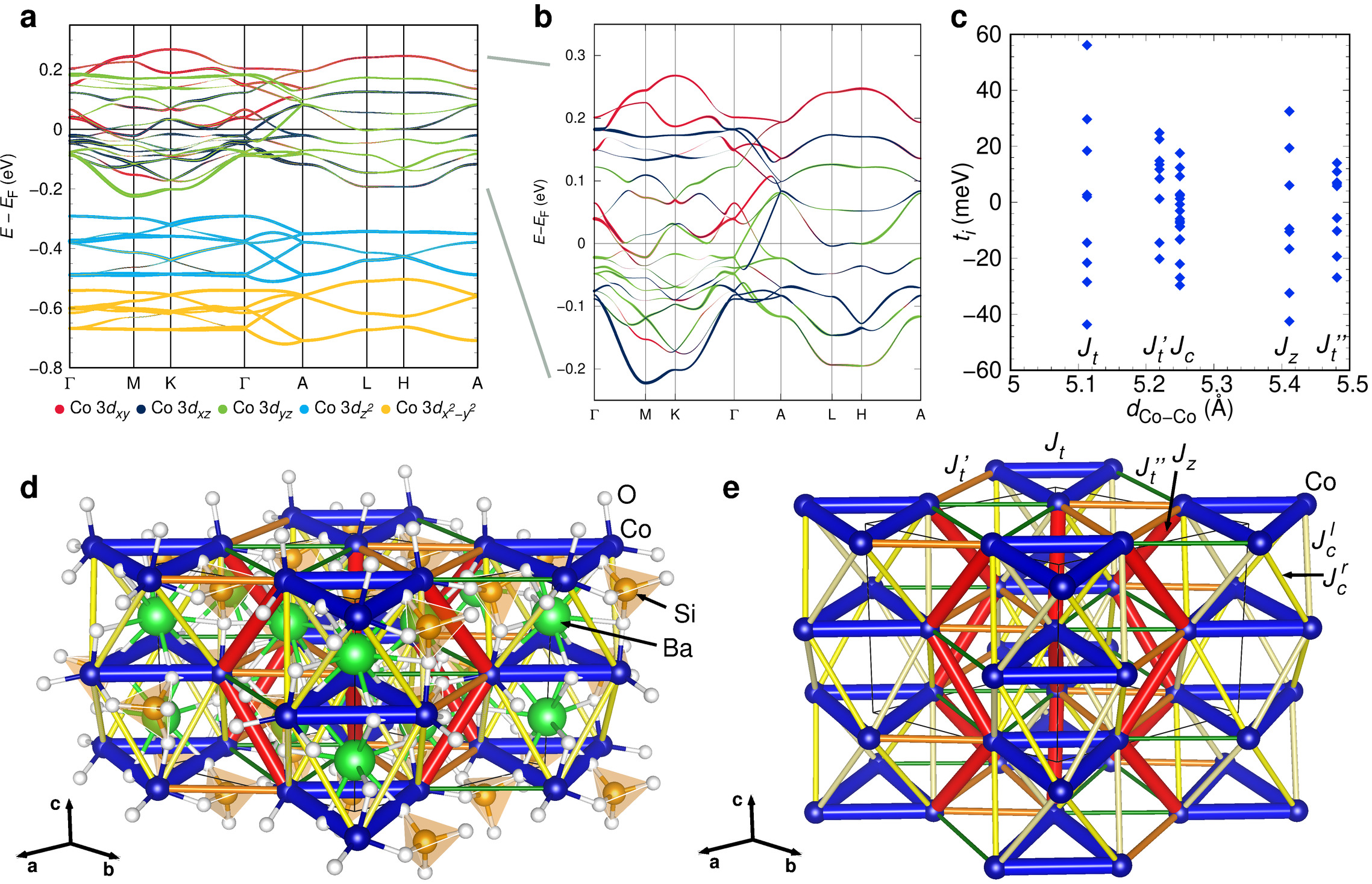}
\caption{ {\bf a} Band structure of {\ba} with $3d$ orbital character of one Co highlighted, calculated with FPLO basis and GGA exchange correlation functional. {\bf b} Tight binding fit of the 18 $t_{2g}$ bands in {\bf a}. Bands and orbital character fits the DFT result perfectly. {\bf c} Tight binding parameters of the five nearest Co-Co distances. Labels indicate to which exchange path the $t_i$ contribute. {\bf d} Structure of {\ba} with the first five Co-Co exchange paths drawn as bonds. Bond cross section is proportional to exchange coupling strength. {\bf e} Co network with exchange paths as in {\bf d}. }\label{fig:elstructure}
\end{figure*}

{\it Electronic structure.-} We calculated the electronic structure of {\ba} using the full potential local orbital (fplo) basis~\cite{fplo} and generalized gradient approximation exchange and correlation functional~\cite{Perdew1996}. 
Fig.~S\ref{fig:elstructure}\,{\bf a} shows the 30 $3d$ bands arising from the six Co$^{2+}$ ions in the unit cell, with $3d$ character of one of the ions highlighted. The local coordinate system is chosen so that the Co-O bonds point into the corners of the cube spanned by the unit vectors. The $e_g$ bands are below the $t_{2g}$ bands and separated from them by a small gap; the $t_{2g}$ bands are half-filled.

{\it Tight binding model.-} In order to get an impression which exchange paths might be important, we use projective Wannier functions~\cite{Eschrig2009} to obtain a tight binding model for the Co $t_{2g}$ electrons. Bands from this model are shown in Fig.~S\ref{fig:elstructure}\,{\bf b} and perfectly match the DFT bands in dispersion and character. Hopping parameters for first five exchange paths are shown in Fig.~S\ref{fig:elstructure}\,{\bf c}. These five paths are shown as bonds in Fig.~S\ref{fig:elstructure}\,{\bf d} and {\bf e}. In-plane couplings $J_t$, $J_t'$ and $J_t''$ form triangles; inter-layer couplings $J_z$ define zigzag chains along $c$, and inter-layer couplings $J_c$ are chiral, with $J_c^r$ defining a right screw and $J_c^l$ a left screw; $J_c^r$ and $J_c^l$ belong to the same distance $d_{\rm Co-Co}=5.250$\,{\AA} but are symmetry inequivalent. 
The TB parameters do not rapidly decrease with increasing distance, and the inter-layer hoppings ($d_{\rm Co-Co}=5.250$\,{\AA} and 5.411\,{\AA} labeled $J_c$ and $J_z$) are not substantially smaller than the three in-plane hoppings. On the contrary, there is a suggestion in these parameters that one in-plane coupling ($J_t$, $d_{\rm Co-Co}=5.113$\,{\AA}) and one inter-layer coupling ($J_z$, $d_{\rm Co-Co}=5.411$\,{\AA}) may play a more prominent role than the other exchange couplings. The conclusion from the tight binding parameters is that it is crucial to calculate all five exchange couplings by energy mapping.

\begin{table}[htb]
\caption{Exchange couplings of {\ba}, calculated for a $\sqrt{2}\times\sqrt{2}\times 1$ unit cell within GGA+U at $J_{\rm H}=0.84$~eV and $4\times 4\times 4$ $k$ points. The last row contains the Co-Co distances which identify the exchange paths. The errors shown are only the statistical error arising from the energy mapping. The interpolated $U=4.41$~eV set of couplings (in bold face) reproduces the (average) experimental Curie-Weiss temperature.}\label{tab:exchange0}
\begin{center}
\begin{tabular}{c|c|c|c|c|c|c}
\hline
\hline
$U$\,(eV)& $J_t$\,(K)&    $J_t'$\,(K)&   $J_c$\,(K)&  $J_z$\,(K)&   $J_t''$\,(K)& $\theta_{\rm CW}$\,(K)\\
\hline
3 & 3.48(1) & 0.74(1) & 0.50(1) & 3.03(1) & 0.27(1) & -21.8 \\
3.5 & 3.11(1) & 0.64(1) & 0.44(1) & 2.64(1) & 0.22(1) & -19.2 \\
4 & 2.80(1) & 0.54(1) & 0.38(1) & 2.32(1) & 0.19(1) & -17.0 \\
{\bf 4.41} & {\bf 2.56(1)} & {\bf 0.49(1)} & {\bf 0.35(1)} & {\bf 2.09(1)} & {\bf 0.16(1)} & {\bf -15.4} \\
4.5 & 2.51(1) & 0.48(1) & 0.35(1) & 2.04(1) & 0.15(1) & -15.1 \\
\hline

$d\,({\rm \AA})$ & 5.113 & 5.220 & 5.250 & 5.411 &  5.481 & \\
\hline
\hline
\end{tabular}
\end{center}
\end{table}

\begin{table*}[htb]
\caption{Exchange couplings of {\ba}, calculated for a $\sqrt{2}\times 1\times\sqrt{2}$ unit cell within GGA+U at $J_{\rm H}=0.84$~eV and $4\times 4\times 4$ $k$ points for the room temperature structure. The last row contains the Co-Co distances which identify the exchange paths. The errors shown are only the statistical error arising from the energy mapping. The interpolated $U=4.37$~eV set of couplings (in bold face) reproduces the (average) experimental Curie-Weiss temperature.}\label{tab:exchange}
\begin{center}
\begin{tabular}{c|c|c|c|c|c|c|c}
\hline
\hline
$U$\,(eV)& $J_t$\,(K)&    $J_t'$\,(K)&   $J_c^r$\,(K)&  $J_c^l$\,(K)&   $J_z$\,(K)&   $J_t''$\,(K)& $\theta_{\rm CW}$\,(K)\\
\hline
3 & 3.45(2) & 0.73(3) & 0.46(3) & 0.62(3) & 3.03(2) & 0.28(3) & -21.6 \\
3.5 & 3.09(2) & 0.62(3) & 0.40(3) & 0.55(3) & 2.65(2) & 0.24(3) & -19.0 \\
4 & 2.77(2) & 0.54(2) & 0.35(2) & 0.50(2) & 2.32(2) & 0.20(2) & -16.8 \\
{\bf 4.37} & {\bf 2.57(2)} & {\bf 0.49(2)} & {\bf 0.32(2)} & {\bf 0.46(2)} & {\bf 2.11(2)} & {\bf 0.17(2)} & {\bf -15.4} \\
4.5 & 2.50(2) & 0.47(2) & 0.31(2) & 0.45(2) & 2.04(2) & 0.16(2) & -14.9 \\
\hline
$d\,({\rm \AA})$ & 5.113 & 5.220 & 5.250 & 5.250 & 5.411 &  5.481 & \\
\hline\hline
\end{tabular}\end{center}
\end{table*}

{\it Energy mapping.-} We now extract the Heisenberg Hamiltonian parameters of {\ba} using the energy mapping technique~\cite{Iqbal2017}. We use two different supercells: With a $\sqrt{2}\times\sqrt{2}\times 1$ cell for which the results are summarized in Table~S\ref{tab:exchange0}, we 
can resolve exchange couplings up to $d_{\rm Co-Co}=9.270$\,{\AA} and convince ourselves that only the first five paths shown in Fig.~S\ref{fig:elstructure}\,{\bf e} are relevant. Using a $\sqrt{2}\times 1\times\sqrt{2}$ supercell, with results reported in Table~S\ref{tab:exchange}, we are able to separate the left-winding and right-winding chiral couplings $J_c^r$ and $J_c^l$. For the first supercell, the 12 independent Co$^{2+}$ moments in $P\,1$ symmetry allow for 460 unique energies of different configurations; we randomly choose 39 of these spin configurations and obtain an excellent fit to the Heisenberg Hamiltonian in the form
\begin{equation}
  H=\sum_{i<j} J_{ij} {\bf S}_i\cdot {\bf S}_j
\end{equation}
Total moments are exact multiples of 3~$\mu_{\rm B}$ as all Co$^{2+}$ 
moments are  $S=3/2$. In Table~S\ref{tab:exchange0}, the values of the $J_i$ are given with
respect to spin operators of length $S=3/2$. Please note that if the
Hamiltonian is written as $\sum_{ij}$ counting every bond twice, the
$J_i$ need to be divided by two. Besides the five couplings shown in Table~S\ref{tab:exchange0}, we find negligibly small longer range couplings $J_6=0.01(1)$\,K ($d_{\rm Co-Co}=7.328$\,{\AA}), $J_7=0.05(1)$\,K ($d_{\rm Co-Co}=7.549$\,{\AA}), $J_9=0.00(1)$\,K ($d_{\rm Co-Co}=8.963$\,{\AA}), $J_{10}=0.00(1)$\,K ($d_{\rm Co-Co}=9.084$\,{\AA}), $J_{11}=0.03(1)$\,K ($d_{\rm Co-Co}=9.126$\,{\AA}), $J_{12}=0.00(1)$\,K ($d_{\rm Co-Co}=9.237$\,{\AA}), $J_{13}=-0.01(1)$\,K ($d_{\rm Co-Co}=9.270$\,{\AA}). The Curie-Weiss temperature estimates are obtained from
\begin{equation}\begin{split}
    \theta_{\rm CW} = &-\frac{2}{3} S (S + 1) \Big(J_1+J_2+2J_3+J_4+J_5+J_6+2J_7+J_9+2J_{10}+3J_{11}+2J_{12}+J_{13} \Big)
\end{split}\end{equation}
where $S=\nicefrac{3}{2}$. The $U$ value is determined by demanding that the couplings reproduce the experimental Curie-Weiss temperature. From fits to the inverse susceptibility, we have $\theta_{\rm CW}=-10(2)$\,K for $H\parallel ab$ and $\theta_{\rm CW}=-26.2(4)$\,K for $H\parallel c$. As approximate energy scale, we use a weighted average of these two values, $\theta_{\rm CW}=-15.4$\,K. The corresponding interpolated set of exchange couplings is given in Table~S\ref{tab:exchange0} in bold face.

For the second supercell ($\sqrt{2}\times 1\times\sqrt{2}$), the 12 independent Co$^{2+}$ moments in $P\,1$ symmetry lead to 195 spin configurations with distinct energies out of which we use 38 for the energy mapping. Besides the six couplings given in Table~S\ref{tab:exchange}, we obtained negligibly small $J_8=0.05(1)$\,K ($d_{\rm Co-Co}=8.683$\,{\AA}) and $J_{14}=0.01(1)$\,K ($d_{\rm Co-Co}=10.076$\,{\AA}). The two chiral couplings $J_c^r$ and $J_c^l$ turn out to be substantially different, with $J_c^r$ 50{\%} larger than $J_c^l$.

{\it Fully relativistic calculations.-} 
We use collinear relativistic DFT calculations to estimate the single ion anisotropy $E_{\rm SIA} = A S_z^2$.
In order to separate $E_{\rm SIA}$ from anisotropic exchange, we calculated total energies for three different spin configurations for the six Co ions in the unit cell: a) FM (ferromagnetic). b) TAFM (one $J_t$ triangle up, one $J_t$ triangle down). c) STRIPY (up-up-down in one $J_t$ triangle, up-down-down in the other). We calculated energies for moments ${\bf m}\parallel {\bf x}$, ${\bf m}\parallel {\bf y}$, and ${\bf m}\parallel {\bf z}$; the first two energies are the same, the third is higher.We use both plain GGA+SO calculations and GGA+SO+U with $U=4$\,eV (the value we know to describe the material correctly). The single ion anisotropy energy estimates are listed in Table~S\ref{tab:SIA}. The fact that the dependence on spin configuration is small (only about 10\% variation) indicates that anisotropic exchange is not strong.

\begin{table}[htb]
\caption{Estimates of single ion anisotropy energies $A$ for three spin configurations. }\label{tab:SIA}
\begin{center}
\begin{tabular}{cccc}
\hline
\hline
         & $A_{\rm FM}$~(K)   & $A_{\rm TAFM}$~(K) & $A_{\rm STRIPY}$~(K) \\
\hline
GGA+SO   & 1.85 & 1.85 & 1.82   \\
GGA+SO+U ($U=4$\,eV) & 2.38 & 2.14 & \\
\hline
\hline       
\end{tabular}\end{center}
\end{table}

\clearpage

\section{Theoretical modeling}
\begin{figure}[h]
\begin{center}
\includegraphics[width=0.7\linewidth]{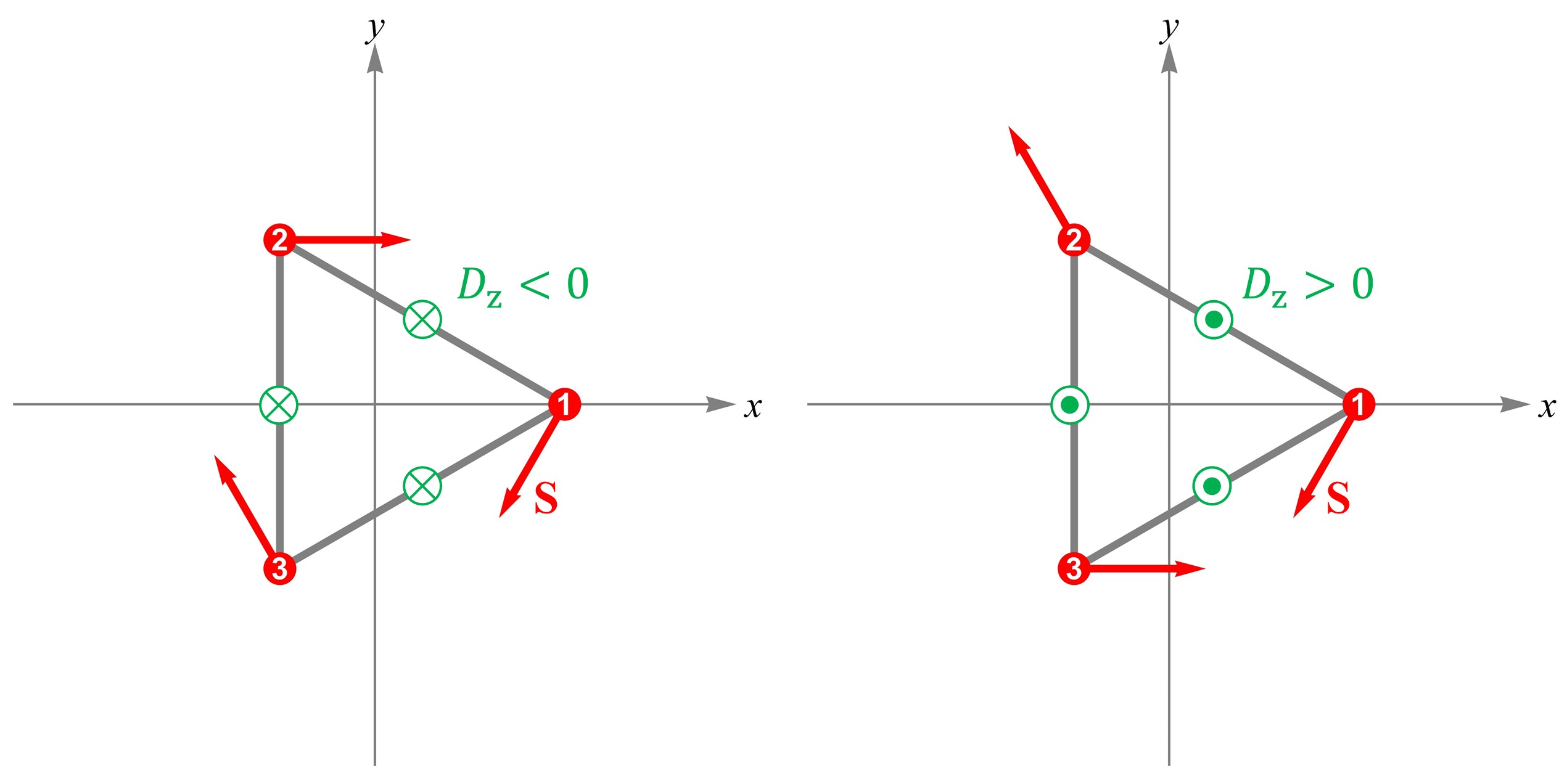}
\end{center}
\caption{Minimal energy configuration for $D_z<0$ (Left) and $D_z>0$ (Right). A uniform rotation applied to all spins does not change energy.}\label{fig:S5}
\end{figure}

Consider one triangle with three spins coupled by the DM interactions that respect the 3-fold rotation symmetry. The total energy is given by $E_\text{DMI}^\text{tot}=({\bf S}_1\times{\bf S}_2)\cdot{\bf D}_3+({\bf S}_2\times{\bf S}_3)\cdot{\bf D}_1+({\bf S}_3\times{\bf S}_1)\cdot{\bf D}_2$, where spins $\{{\bf S}_i,\,i=1,2,3\}$ are numbered counterclockwise on the triangle, and ${\bf D}_i$ is the DM vector on the opposite edge of ${\bf S}_i$. The energy associated with the out-of-plane component of the DM vectors is $E_\text{DMI}^\text{z}=D_z\hat{z}\cdot\left({\bf S}_1\times{\bf S}_2+{\bf S}_2\times{\bf S}_3+{\bf S}_3\times{\bf S}_1\right)\equiv D_z\hat{\bf z}\cdot\boldsymbol\epsilon$, where $\boldsymbol \epsilon$ is the vector spin chirality. This term always favors coplanar spin configurations, for which $|\hat{\bf z}\cdot\boldsymbol\epsilon|$ is maximized (Fig.~S\ref{fig:S5}). 

\begin{figure}[h!]
\centering
  \includegraphics[width=0.9\textwidth]{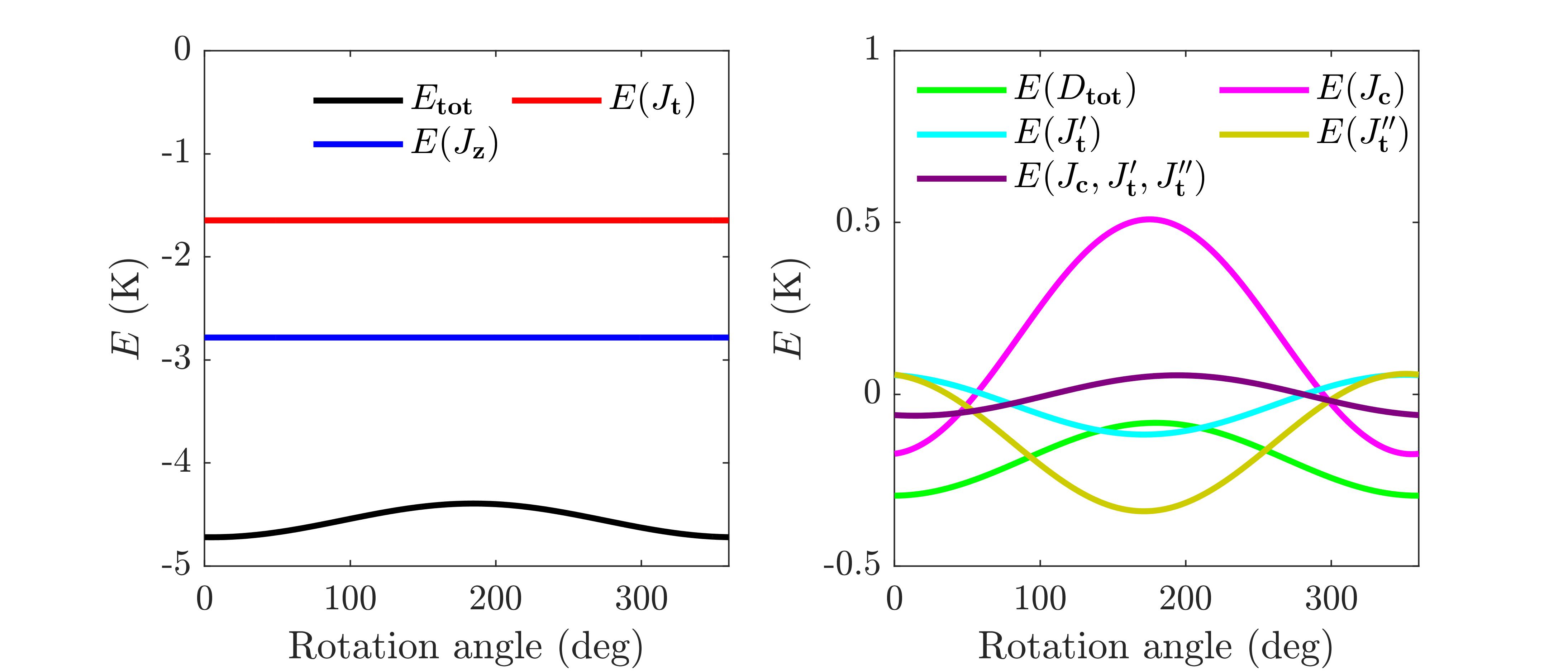}
\caption{Energy per spin for each of the terms in the Hamiltonian as a function of rotation angle. }\label{Fig:rotation}
\end{figure}

Focusing on spin configurations with the 3-fold rotation symmetry, we derive the total DMI energy with a spin and the DM vector on its opposite edge written in the same local frame, ${\bf V}_i = V_z\hat{\bf z}+V_\perp(-\hat{\bf r}_i)+V_\parallel\hat{\bf z}\times(-\hat{\bf r}_i)$ where $V\equiv S$ or $D$ [Fig.~3a], and $\hat{\bf r}_i$ is the unit vector from the center of the triangle pointing to site $i$. The result is 
\begin{equation}
E_\text{DMI}^\text{tot}=\frac{3}{2} \sqrt{3}D_z \left(S_\parallel^2+S_\perp^2\right)+3 \sqrt{3} S_z (D_\parallel S_\parallel+D_\perp S_\perp)
\equiv \frac{3}{2} \sqrt{3}D_z |{\bf S}_{xy}|^2+3 \sqrt{3} S_z {\bf S}_{xy}\cdot{\bf D}_{xy}\,,
\end{equation}
where ${\bf S}_{xy} = (S_\perp,S_\parallel)$ with the $\!{\bf S}_i$ component parallel to radius-vector $\!{\bf r}_i$ as $S_\parallel$  and that perpendicular to it $S_\perp$, and ${\bf D}_{xy} = (D_\perp,D_\parallel)$. 
Other relevant physical quantities can be calculated: toroidal moment $\textbf{t}\!=\!\sum_i {\bf r}_i \times {\bf S}_i = 3S_\perp\hat{\bf z}$; 
vector spin chirality ${\boldsymbol\epsilon}\!=\!{\bf S}_1 \times {\bf S}_2 + {\bf S}_2 \times {\bf S}_3 + {\bf S}_3 \times {\bf S}_1=\frac{3\sqrt{3}}{2}S_{xy}^2\hat{\bf z}$; 
scalar spin chirality $\kappa\!=\!({\bf S}_1 \times {\bf S}_2) \cdot {\bf S}_3=|\boldsymbol\epsilon|S_z$.

To understand the role played by the subleading interactions, we start with the energy-minimized ferritoroidal structure in zero field, continuously rotate along the c-axis all spins on the sublattice with toroidal moment opposite to the net moment and plot the energy of each term in the Hamiltonian as a function of rotation angle in Fig.~S\ref{Fig:rotation}. The left panel shows the total energy (black), the energy of $J_{t}$ bonds (red) and that of $J_{z}$ (blue). Since the global rotation occurs within one of $J_{t}$-$J_{z}$ sublattices, the energy of these two bonds stays the same. 
The right panel of Fig.~S\ref{Fig:rotation} shows the energy of all subleading interactions. It is evident that the $J_{c}$ term favors the ferritoroidal state, while $J_{t}'$ and $J_{t}''$ favors the ferrotoroidal one.

\bibliographystyle{naturemag}